\documentclass{article}
\UseRawInputEncoding
\usepackage{graphicx} 
\usepackage[numbers,sort&compress]{natbib} 
\usepackage{hyperref}
\usepackage{float}
\usepackage{listings}
\usepackage{colortbl,xcolor}
\usepackage{amsmath}
\usepackage{booktabs}
\title{Tech Report: Divide and Conquer 3D Real-Time Reconstruction for Improved IGS}
\author{Yicheng Zhu}
\date{December 2024}

\begin{document}

\maketitle

\begin{abstract}

Tracking surgical modifications based on endoscopic videos is technically feasible and of great clinical advantages; however, it still remains challenging. 
This report presents a modular pipeline to divide and conquer the clinical challenges in the process. The pipeline integrates frame selection, depth estimation, and 3D reconstruction components, allowing for flexibility and adaptability in incorporating new methods. Recent advancements, including the integration of Depth-Anything V2 and EndoDAC for depth estimation, as well as improvements in the Iterative Closest Point (ICP) alignment process, are detailed. Experiments conducted on the Hamlyn dataset demonstrate the effectiveness of the integrated methods. System capability and limitations are both discussed.
\end{abstract}

\section{Introduction}

\paragraph{Background}
Image-guided surgery (IGS) technology, by displaying surgical instruments on CT/MRI images in real-time, grants surgeons the ability to see instruments inside a patient and to see pathology that is not visible in an endoscope, making endoscopic approaches possible in some procedures \cite{transorbital,snyderman2009limits}.  A preoperative CT/MRI scan is used as a reference map to visually provide the position information to the surgeon, showing the position of the surgical instrument tip inside a patient and lesions that are not visible in endoscopic images \cite{igs-intro}.

Endoscopic Sinus and Skull Base Surgery (ESSBS) is a procedure that covers the wide sinus and skull base area. The anatomical structures are similar and surrounded by abundant amounts of nerves and blood vessels, making it challenging to ensure surgical completeness. The revision rate of ESSBS is close to 28\% \cite{gpskbcc05,kpcstgz11,swsms11}.

\paragraph{Problem Statement and Motivation}
However, the reference map of the patient is static and does not account for tissue removal, deformation, or other intraoperative changes. In order to update tissue removal and deformation, intraoperative CT/MRI scans need to be performed \cite{igs-intro}, which expose the patient to an extra amount of radiation and anesthesia.

Tracking surgical modifications in real-time with endoscopic videos and instrument motion will revolutionize IGS as it will not only improve navigation precision and reliability but also potentially extend IGS to soft-tissue dominant endoscopic surgeries.

The motion-based surgical modification tracking is rarely studied because IGS motion data not only suffers low sample rate and occlusion but also has unpredictable tracking failures\cite{JNSB18RelativeMotion, IJCARS22Virtual, ICRA21SkillAssessment,harbison2016objective}.

Tracking surgical modifications with endoscopic images in real-time, on the other hand, is an appealing possible solution and has been extensively studied over the past two decades\cite{sinusRegistration,endoscopicSinus}. {However, the classic visual reconstruction methods, including the geometric models and the viewing angle difference-based algorithms (often known as Structure from Motion (SfM)\cite{sfm}), the 3-Dimensional geometry, and the light source changes based algorithms (often known as Shape-from Shading (SfS)\cite{sfs2}), and the consistency of motions and observations (often known as Visual Simultaneous Localization and Mapping (vSLAM) \cite{monoslam}), can not maintain convergence under the adverse conditions in ESSBS, such as bleeding, occlusion, reflection, over-exposure, etc. The recent deep-learning reconstruction methods address various adverse conditions independently \cite{Sensors24Recon}; however, they lack methods and data to address the dimensionality explosion problem caused by the presence of all adverse conditions in ESSBS, and \emph{representative state of art deep learning algorithms diverge within a few minutes in operating rooms}\cite{endodepth}. }

\paragraph{Objective}
The primary objective of this current pipeline is to establish a foundational framework for 3D reconstruction in endoscopic surgeries. Recent advancements have been made to integrate new depth estimation methods and improve the 3D reconstruction process. The aims are:

\begin{enumerate}
    \item \textbf{Preprocessing}: Handle different kinds of inputs, including video files, and extract frames as needed. 
    \item \textbf{Frame Selection}: Implement and refine frame selection methods to filter out informative frames, reducing noise and eliminating irrelevant or low-quality frames. Currently implemented frame selection methods include
    \begin{itemize}
        \item HyperIQA \cite{Su_2020_CVPR}, choosing frames based on their quality assessment score. 
        \item R-channel intensity, choosing frames based on their average r-channel intensity.
    \end{itemize}
    \item \textbf{Depth Estimation}: Integrate new depth estimation methods such as Depth-Anything V1, V2 \cite{yang_depth_2024, yang_depth_2024-1} and EndoDAC\cite{cui_endodac_2024}, and implement necessary post-processing steps.
    \item \textbf{3D Reconstruction}: Implement the point cloud generation using ICP alignment techniques, dynamic thresholding, and artifact removal methods.
\end{enumerate}

\section{Methodology}

Modern visual reconstruction algorithms reach sub-pixel accuracy in ideal laboratory environments but fail to maintain performance in complex real-world environments \cite{Sensors24Recon}).  This is because visual reconstruction requires continuous tracking of scene changes, but various problems deteriorate performance and eventually cause failures\cite{sfm,sfs2,monoslam}. Endoscopic surgeries have extremely challenging scenes, and even humans can lose track. To overcome the problem, we will use a DCL framework that divides and conquers the visual tracking problem in ESSBS, detects and isolates adverse factors, excludes images containing unrecoverable errors from the reconstruction process, compensates for the impacts of adverse factors, reduces the uncertain information contained in an image, and image quality improvement and joint multi-image reconstruction track soft tissue deformation and improve reconstruction accuracy.

\paragraph{Pipeline Overview}
The pipeline code is implemented in Python, with a modular structure and extensive in-line documentation. It is optimized for IDEs like PyCharm and VSCode by utilizing region-based code folding. The repository is available at: \url{https://github.com/Mayonezyck/pipeline}. Users can refer to the \texttt{README.md} for detailed installation and configuration instructions. The pipeline is structured around a configuration YAML file (\texttt{CONFIG.yaml}), which enables flexible and easily modifiable parameters for each stage of the pipeline.

\subsection{Frame Selection}
The pipeline begins with a frame selection step that filters out uninformative or low-quality frames from the input endoscopic video or image folder. This step can run multiple selection schemes in sequence, controlled by parameters in \texttt{CONFIG.yaml}.

\begin{enumerate}
    \item \textbf{HyperIQA}: The pipeline uses HyperIQA \cite{Su_2020_CVPR}, a blind image quality assessment model. For each frame, it estimates a quality score. Frames are then thresholded based on the IQA score. The code implementation (in \texttt{frameSelect/hyperIQA/hyperIQA.py}) loads a pre trained HyperIQA model, runs inference on each frame (with random crops for robustness), and returns a mean quality score. 

    \begin{equation}
    \text{QualityScore} = \frac{1}{N} \sum_{i=1}^{N} f_{\text{HyperIQA}}(I)  
    \end{equation}

    where $f_{\text{HyperIQA}}$ is the HyperIQA inference function, $I$ is the input image, and $N$ is the number of random crops used.

    A threshold $T_{\text{IQA}}$ is then applied:
    \[
    \text{SelectFrame}(I) = 
    \begin{cases}
    1 & \text{if } \text{QualityScore}(I) > T_{\text{IQA}}\\
    0 & \text{otherwise}
    \end{cases}
    \]

    \item \textbf{R\_Channel}: A heuristic filter (\texttt{frameSelect/rchannel.py}) that uses the mean intensity of the red channel to select frames. If the red channel average intensity $R_{\text{mean}}$ is above a user-specified threshold $T_{R}$, the frame is considered to be from the inside environment and not an out-of-patient shot. Formally:
    \[
    \text{SelectFrame}(I) = 
    \begin{cases}
    1 & \text{if } R_{\text{mean}}(I) > T_{R}\\
    0 & \text{otherwise}
    \end{cases}
    \]

\end{enumerate}

At the end of the selection process, a curated list of frames is produced, and a temporary folder is created to store these selected frames. This step ensures that downstream operations only work on frames of adequate quality and relevance.

\subsection{Depth Estimation}

Depth estimation is crucial for 3D reconstruction. The pipeline supports multiple depth estimation algorithms, with configurations selected through \texttt{CONFIG.yaml}. The current codebase implements:

\begin{itemize}
    \item \textbf{Depth-Anything:v2} \cite{yang_depth_2024}: This serves as the primary depth estimation method. The code (in \texttt{depthEstimate/Depth-Anything-V2}) runs a command-line interface to generate depth maps for each selected frame. Each frame $I$ is mapped to a depth image $D$ through:
    \[
    D = f_{\text{Depth-Anything:v2}}(I; \theta)
    \]
    Where $\theta$ are the pre-trained model parameters.

    \item \textbf{EndoDAC} \cite{cui_endodac_2024}: A state-of-the-art endoscopic depth estimation model trained on the SCARED dataset. Implemented in \texttt{depthEstimate/EndoDAC}, it produces disparity maps that are converted to depth using camera parameters. Postprocessing steps are discussed later.

    \begin{figure}[H]
        \centering
        \includegraphics[width=0.7\linewidth]{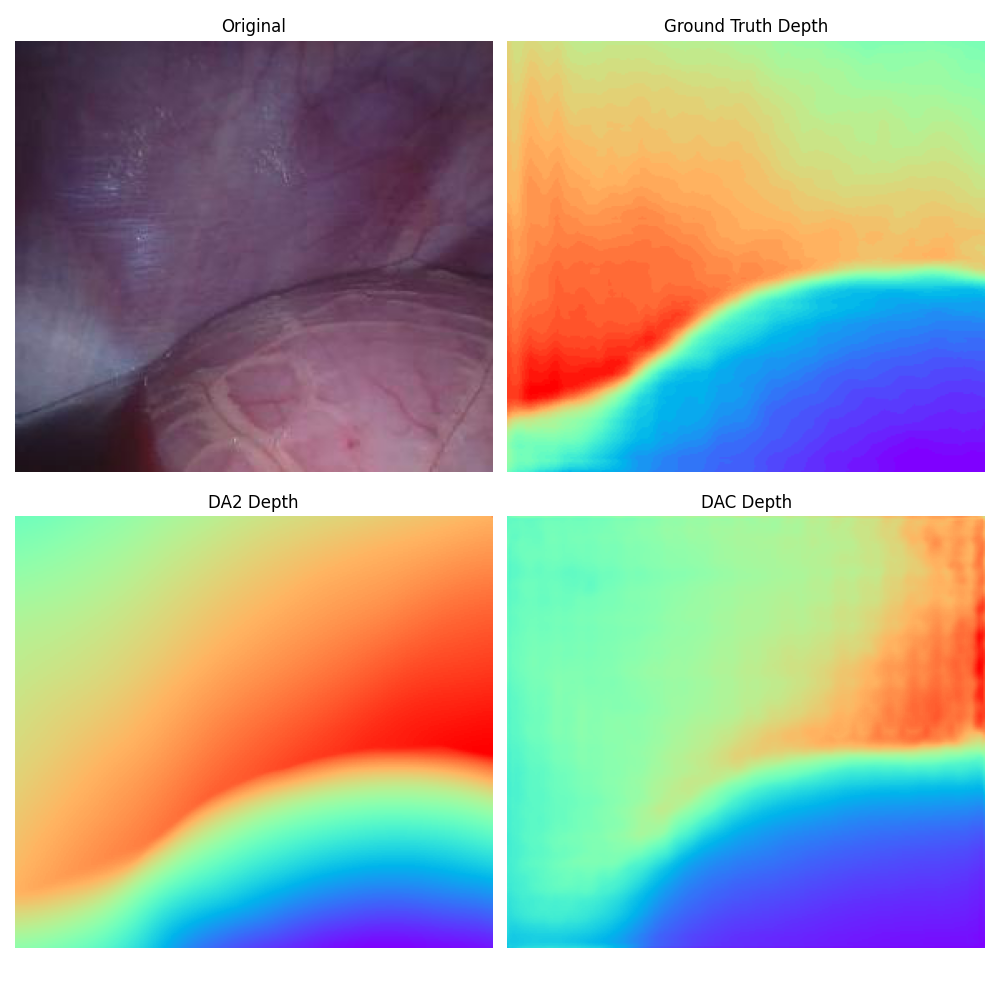}
        \caption{Visualization comparing Depth-Anything V2 and EndoDAC results on a sample frame.[The first frame of test22 from Hamlyn.]}
        \label{fig:dav2demo}
    \end{figure}
    \begin{figure}[H]
        \centering
        \includegraphics[width=0.7\linewidth]{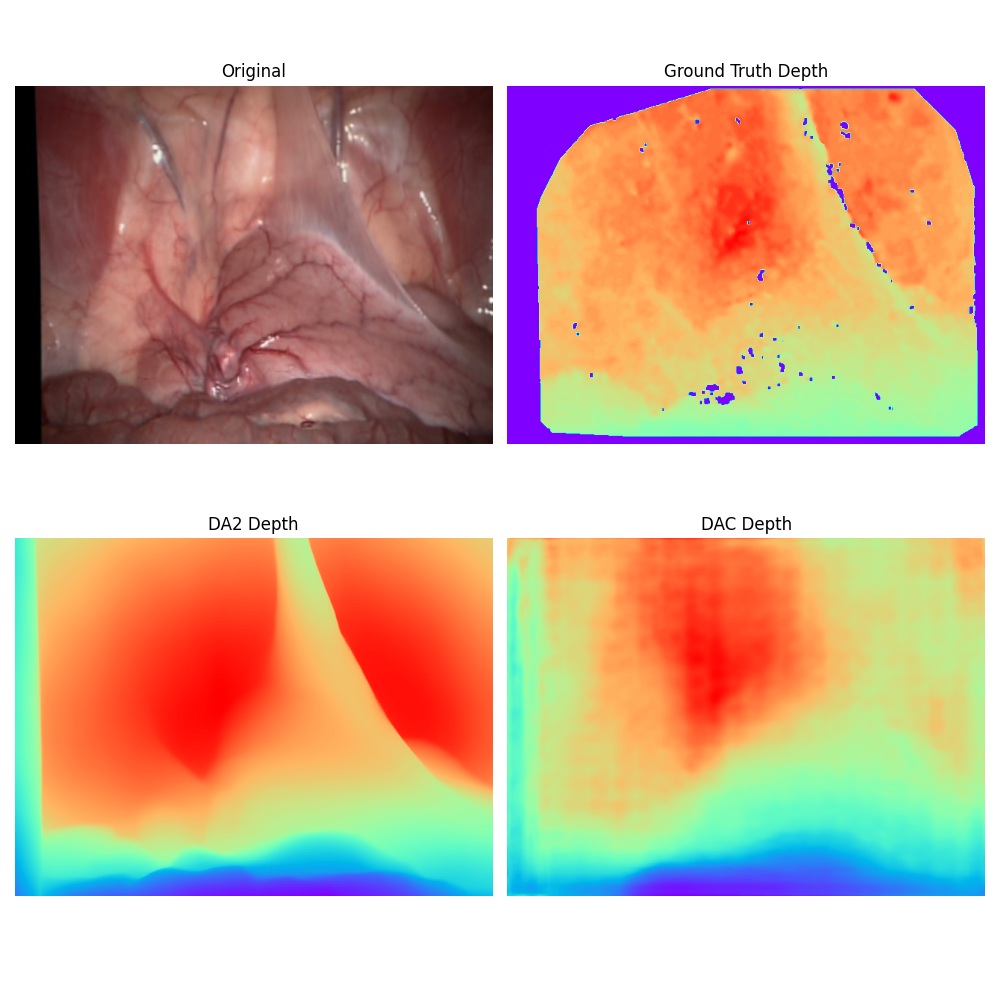}
        \caption{Visualization comparing Depth-Anything V2 and EndoDAC results on a sample frame.[The first frame of rectified01 from Hamlyn.]}
        \label{fig:dav2demo2}
    \end{figure}
\end{itemize}

After prediction, a universal mask is applied to remove black borders or invalid pixels from all depth maps. This mask is generated by comparing pixel intensities across all input frames (see the \texttt{generate\_universal\_mask} function in the ICP code). Additionally, post-processing steps can handle the normalization and scaling of depth maps.

\subsection{3D Reconstruction}

The final step merges depth information from multiple frames to form a unified 3D point cloud. The pipeline implements an Iterative Closest Point (ICP)\cite{arun_least-squares_1987, 121791} based reconstruction method. Two modes are supported:

\begin{itemize}
    \item \textbf{ICP with Global Alignment}: Each new frame's depth map, after conversion to a point cloud, is aligned to the global map. The transformation $T$ that minimizes point-to-point distances is estimated iteratively:
    \[
    \min_{R, t} \sum_{i} \| R \mathbf{x}^B_i + t - \mathbf{x}^A_{\text{closest}(i)} \|^2
    \]
    Where $(R,t)$ is the rigid transformation (rotation and translation) from the source point cloud $B$ to the reference point cloud $A$, the alignment is done repeatedly, integrating each new frame into the global model.

    \item \textbf{ICP with Neighbor Alignment}: Instead of aligning each frame to the global map directly, frames may be aligned to their temporal neighbors to reduce drift. This approach can improve local consistency.

\end{itemize}

The ICP variants are implemented in \texttt{reconstruction/icp\_neighbor.py} and \texttt{reconstruction/icp.py}, allowing switching between a custom ICP routine and Open3D's built-in ICP method \cite{zhou_open3d_2018}. The custom ICP uses a KD-tree search (via \texttt{scipy.spatial.cKDTree} \cite{virtanen_scipy_2019}) to find closest point correspondences and applies dynamic thresholding to filter out non-overlapping regions or outliers.

\paragraph{Dynamic Thresholding Mechanism}
Dynamic thresholds are applied on distances between correspondences to focus on well-overlapping points:
\[
\text{mask}(i) = 
\begin{cases}
1 & \text{if } d(i) < T_d \\
0 & \text{otherwise}
\end{cases}
\]

Threshold $T_d$ can be constant or adaptively chosen based on statistical measures (mean, median, standard deviation, or percentile of observed distances). The code implements adaptive thresholding to exclude points that lie beyond a certain distance, improving convergence and reducing the influence of outliers.

\begin{figure}[H]
    \centering
    \includegraphics[width=0.4\linewidth]{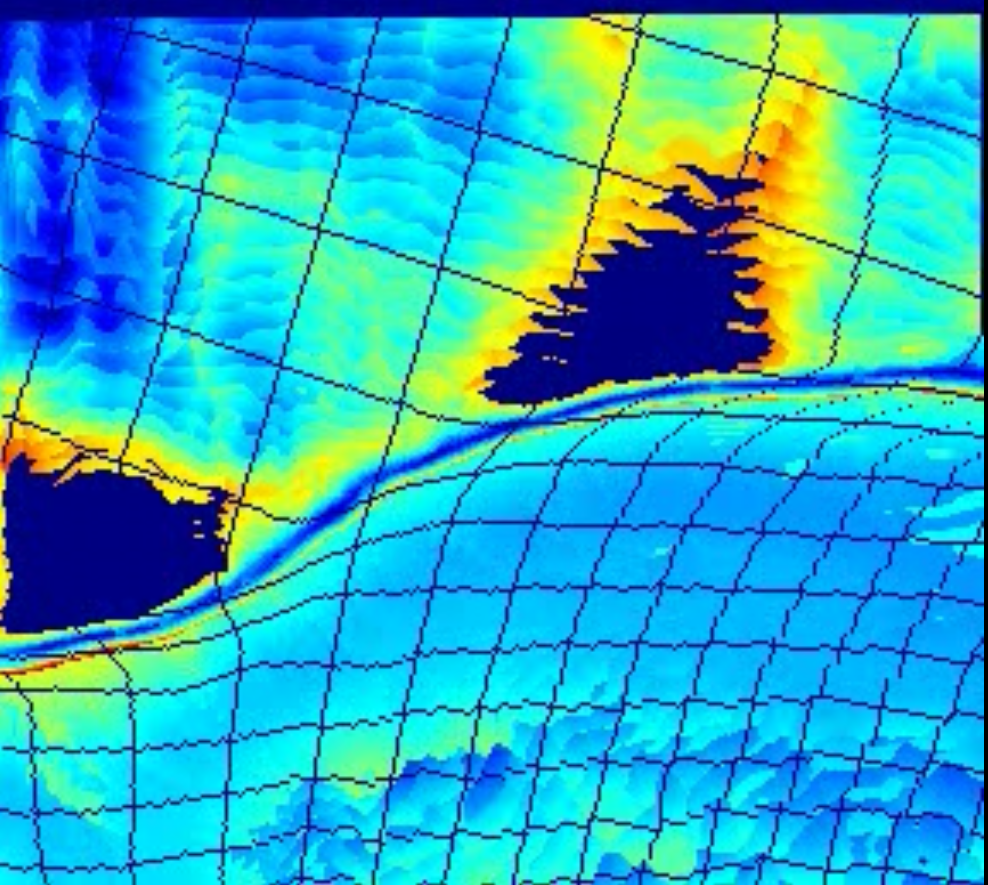}
    \includegraphics[width=0.4\linewidth]{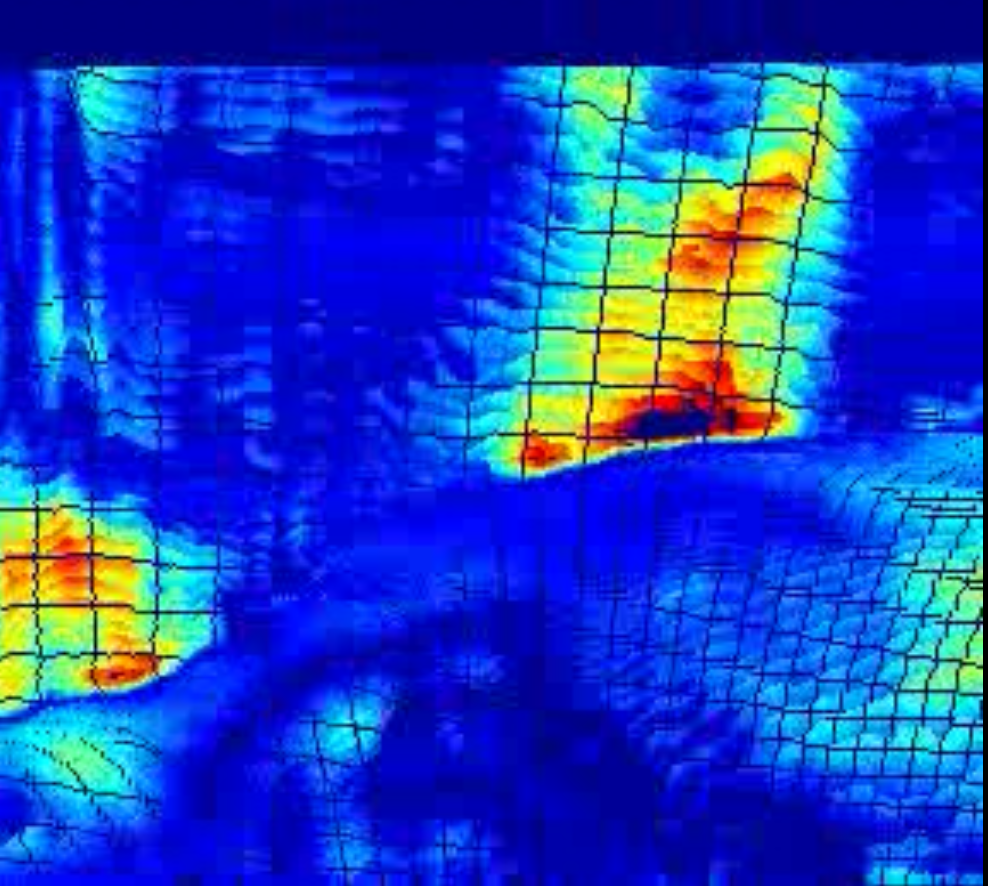}
    \caption{Heatmap of the per-pixel alignment error at the first (left) and tenth (right) ICP iteration. Darker areas indicate masked-out regions, while dark blue areas represent smaller errors.}
    \label{fig:heatmap_constant}
\end{figure}

\paragraph{Visualization and Error Analysis}
Various visualization tools are integrated to facilitate debugging and parameter tuning. The pipeline can generate heatmaps representing alignment errors per pixel. In Figure \ref{fig:heatmap_constant}, we show an illustration of such error heatmaps for the first and tenth iterations of ICP:
\[
\text{Error Heatmap}(x, y) =
\begin{cases} 
\text{Distance}(\text{Point}_1, \text{Point}_2), & \text{if a matching point is found,} \\
0, & \text{if no matching point is found.}
\end{cases}
\]

The code also outputs diagnostic plots, logs intermediate transformations, and can save intermediate point clouds for manual inspection.

\subsection{Technical Implementation Details from the Code}
Below are some additional technical details tied directly to the code structure:

\begin{itemize}
    \item \textbf{Configuration Loading (main.py)}: The pipeline reads \texttt{CONFIG.yaml} via \texttt{dataloader/yaml\_handler.py}, setting parameters like \texttt{DATA\_PATH}, \texttt{SELECT\_SCHEME}, and \texttt{DEPTH\_SCHEME}. 

    \item \textbf{Frame Extraction and Temporary Folder Management (main.py)}: The \texttt{frame\_handler} and \texttt{temp\_folder} modules manage frames. Selected frames are copied to a temporary directory \texttt{/temp} for standardized input to depth estimation scripts.

    \item \textbf{Depth Estimation Scripts (estimate.py)}: Depending on the chosen method (e.g., \texttt{depth-anything:v2}), the code constructs command-line arguments and invokes external Python scripts (e.g., \texttt{run.py}) in the corresponding subfolder. Parameters such as \texttt{GRAYSCALE} or \texttt{PRED\_ONLY} are appended based on the YAML configuration.

    \item \textbf{Postprocessing (postprocessing.py)}: 

After obtaining the raw depth or disparity maps, a postprocessing stage (as implemented in \texttt{postProcessing/postprocess.py}) refines the results. Consider the case where EndoDAC predictions output disparity maps \texttt{\_disp.npy}. The pipeline:

1. \textbf{Resizes} the disparity map to match the original frame's resolution $(W, H)$.  
   Given the original image size $(W, H)$:
   \[
   \tilde{\text{disp}} = \text{Resize}(\text{disp}, W, H)
   \]

2. \textbf{Converts Disparity to Depth} using the provided $f_x$ and baseline $b$:
   \[
   D(x,y) = \frac{f_x \cdot b}{\tilde{\text{disp}}(x,y)}
   \]

3. \textbf{Normalization}: The resulting depth map is then normalized to an 8-bit scale (0--255) for visualization and storage. Let $D_{\text{min}}$ and $D_{\text{max}}$ be the minimum and maximum values in $D$, respectively:
   \[
   D_{\text{norm}}(x,y) = \frac{D(x,y) - D_{\text{min}}}{D_{\text{max}} - D_{\text{min}}} \cdot 255
   \]

   This produces a grayscale depth image suitable for immediate inspection.

An example workflow for EndoDAC postprocessing from the code snippet is:

\begin{enumerate}
    \item Locate the \texttt{\_disp.npy} files in the prediction output directory.
    \item For each \texttt{\_disp.npy} file, load the disparity map and resize it to the original image dimensions.
    \item Compute depth from disparity.
    \item Normalize $D$ and save as a \texttt{\_depth.png} image.
\end{enumerate}

    \item \textbf{Custom ICP Implementation (icp\_neighbor.py and icp.py)}: The ICP routine uses SVD-based point set alignment. Given correspondences, the rotation $R$ and translation $t$ are computed from the covariance matrix $H$ : 
    \[
    H = \sum_i (\mathbf{x}_i^B - \bar{\mathbf{x}}^B)(\mathbf{x}_{\text{closest}(i)}^A - \bar{\mathbf{x}}^A)^\top
    \]
    \[
    U,S,V^T = \text{SVD}(H), \quad R = VU^T, \quad t = \bar{\mathbf{x}}^A - R \bar{\mathbf{x}}^B
    \]

    To handle outliers points with distances above a dynamic threshold are excluded before computing $R$ and $t$.

    \item \textbf{Point Cloud Merging and K-D Tree Filtering}: In the merging step, a KD-tree is built on the global map. Points from the new frame that lie beyond certain distance thresholds are considered to represent new areas and are appended to the global map, while duplicates or erroneous points are discarded.

    \item \textbf{Evaluation and Visualization}: The code provides functionalities in \texttt{evaluation/evaluate.py} and \texttt{visualize/outputvisual.py} to measure reconstruction quality against ground truth and output intermediate results, if available.

\end{itemize}

\section{Experiments and Evaluation}
We conducted three main experiments to evaluate key components of our pipeline: 

\begin{itemize}
    \item depth-estimation and its quantitative analysis
    \item ICP threshold tuning
    \item end-to-end 3D reconstruction with qualitative assessment.
\end{itemize}
\paragraph{The choice of dataset}
The three implemented methods were tested on two datasets, the 'test22' and 'rectified01', both from the Hamlyn dataset. The 'test22' is chosen because its structure makes it easy for humans to qualitatively judge the performance of the process. 'Rectified01' is chosen because it's a dataset where the camera is placed almost vertically to the surface, so the ground truth is likely to be precise.

\paragraph{Quantative study of depth estimation methods}

Standard evaluation metrics were used to assess the performance of the depth estimation methods, including RMSE, MAE, Square relative error, $\delta$ accuracy($\delta < 1.25$), SSIM \cite{1284395}, and Log RMSE. The establishment of those evaluation protocols follows the standard stated in Eigen et al.'s work \cite{eigen_depth_nodate}.
\paragraph{Metrics}
\begin{itemize}
    \item Root Mean Squared Error (RMSE):
    \begin{equation}
        RMSE = \sqrt{\frac{1}{N} \sum_{i=1}^{N} (d_i^{\text{pred}} - d_i^{\text{gt}})^2}
    \end{equation}
    \item Mean Abolute Error (MAE):
    \begin{equation}
        MAE = \frac{1}{N} \sum_{i=1}^{N} |d_i^{\text{pred}} - d_i^{\text{gt}}|
    \end{equation}
    \item Squared Relative Error (Sq Rel):
    \begin{equation}
        SqRel = \frac{1}{N} \sum_{i=1}^{N} \frac{(d_i^{\text{pred}} - d_i^{\text{gt}})^2}{d_i^{\text{gt}}}
    \end{equation}
    \item Delta Accuracy (Threshold Accuracy):

    For a threshold $\delta$:
    \begin{equation}
        \delta = \frac{\max(d_i^{\text{pred}}, d_i^{\text{gt}})}{\min(d_i^{\text{pred}}, d_i^{\text{gt}})}
    \end{equation}
    The accuracy is then calculated as the proportion of pixels where $\delta$ is less than a certain value, in our case (threshold = 1.25)
    \begin{equation}
        \text{Delta Accuracy} = \frac{1}{N} \sum_{i=1}^{N} [\delta < 1.25]
    \end{equation}
    \item Structural Similarity Index (SSIM):
    \begin{equation}
        SSIM(x, y) = \frac{(2\mu_x \mu_y + C_1)(2\sigma_{xy} + C_2)}{(\mu_x^2 + \mu_y^2 + C_1)(\sigma_x^2 + \sigma_y^2 + C_2)}
    \end{equation}
    where $\mu_x$ and $\mu_y$ are the means of the predicted and ground truth depth maps, $\sigma_x^2$ and $\sigma_y^2$ are the variances, and $\sigma_xy$ is the covariance. $C_1$ and $C_2$ are constants.
    The code used for SSIM is imported from skimage.metrics \cite{van_der_walt_scikit-image_2014}. The data range is set to be:
    \begin{equation}
        \text{data range} = max(max(d^{pred}, max(d^{gt}))
    \end{equation}
    \item Log RMSE:
    \begin{equation}
        \text{Log RMSE} = \sqrt{\frac{1}{N} \sum_{i=1}^{N} \left( \log(d_i^{\text{pred}}) - \log(d_i^{\text{gt}}) \right)^2}
    \end{equation}
\end{itemize}

\paragraph{Finding the best ICP Threshold} When performing ICP, a threshold $T$ is set to help identify the actual corresponding points against outliers, e.g., the points that are new features that don't appear in the overlapping pov. If the threshold is too high, all the points will be used for alignment, leading to an alignment that is heavily affected by the outliers; on the other hand, if the threshold is not high enough, the alignment will not be thorough and is likely to converge at a local minimum. This experiment focuses on trying different schemes of T threshold choosing. Maximum iteration is set to be 40. The scheme of constant value, 90th percentile, factored mean of error, factored median of error, linear interpolation, 80\% maximum distance, and mean plus two standard deviations were compared.

\paragraph{Qualitative test for 3D reconstruction} For both datasets: test22 and recitifed01, depth-anything:v2 was chosen to generate the depth estimation for 3D reconstruction. All frames were taken -- no frame selection were used -- and the output point cloud is kept at frame 1 and frame 30, to compare with the point cloud reconstructed using the same method but having ground truth depth images as input. 

\paragraph{About the testing of the frame selection module}
The frame selection module was not tested as part of the current pipeline evaluation. The datasets we are currently using are well-curated, with off-site frames already cropped out and no presence of liquid interference or low-quality frames. This level of preprocessing negates the immediate necessity of frame selection. However, this module will be rigorously tested in the future when our custom dataset, which includes more varied and challenging conditions, is put into use. A quantitative study of the precision of reconstruction will be performed in the future after the reconstruction methods are optimized, and the performance will be compared between different frame selection schemes. 

\paragraph{Adjustments Made}
Scaling factors were applied to align the predicted depth maps with the ground truth, for Depth-Anything V2, post-processing involved inverting the depth maps and adjusting the color schemes to ensure correct visualization and comparison.

\section{Results}
\subsection{Depth Estimation Performance evaluation}

\paragraph{Quantitative Results}
Experiments showed that Depth-Anything V2(DA:v2) provided better depth estimation results compared to EndoDAC(DAC) and Depth-Anything V1(DA:v1). Metrics such as RMSE and MAE were consistently lower for Depth-Anything V2 across different datasets (Tables \ref{tab:metrics_comparison22}, \ref{tab:metrics_comparison01}).

\begin{table}[h!]
\centering
\caption{Performance Metrics for the three methods on Test22}
\begin{tabular}{@{}lccc@{}}
\toprule
\textbf{Metric}               & \textbf{DAC} & \textbf{DA:v1} & \textbf{DA:v2} \\ \midrule
Mean RMSE                     & 57.1918                 & 24.4134                           & 19.7962                           \\
Mean MAE                      & 42.8775                 & 19.1280                           & 15.0984                           \\
Mean Sq Rel                   & 63.1228                 & 10.7617                           & 8.0148                            \\
Mean $\delta$ Accuracy        & 0.3254                  & 0.5640                            & 0.6619                            \\
Mean SSIM                     & 0.7211                  & 0.8531                            & 0.8771                            \\
Mean Log RMSE                 & 0.7350                  & 0.5599                            & 0.4734                            \\ \bottomrule
\end{tabular}
\label{tab:metrics_comparison22}
\end{table}

\begin{table}[h!]
\centering
\caption{Performance Metrics for the three methods on Rectified01}
\begin{tabular}{@{}lccc@{}}
\toprule
\textbf{Metric}               & \textbf{DAC} & \textbf{DA:v1} & \textbf{DA:v2} \\ \midrule
Mean RMSE                     & 51.8964                 & 26.5211                           & 22.1255                           \\
Mean MAE                      & 35.5692                 & 15.3696                           & 11.6590                           \\
Mean Sq Rel                   & 33.0954                 & 6.4422                            & 3.8836                            \\
Mean $\delta$ Accuracy        & 0.3821                  & 0.6839                            & 0.7815                            \\
Mean SSIM                     & 0.5848                  & 0.6489                            & 0.6611                            \\
Mean Log RMSE                 & 0.6520                  & 0.3826                            & 0.2978                            \\ \bottomrule
\end{tabular}
\label{tab:metrics_comparison01}
\end{table}

\paragraph{Anomalies and Observations}
In some cases, error spikes were observed at certain frames, likely due to erroneous depth maps(shown below in Figure. \ref{fig:errorcomp}) or inconsistencies in the dataset. These anomalies highlighted the need for improved data handling and possibly more robust depth estimation methods.
\begin{figure}[H]
        \centering
        \includegraphics[width=0.7\linewidth]{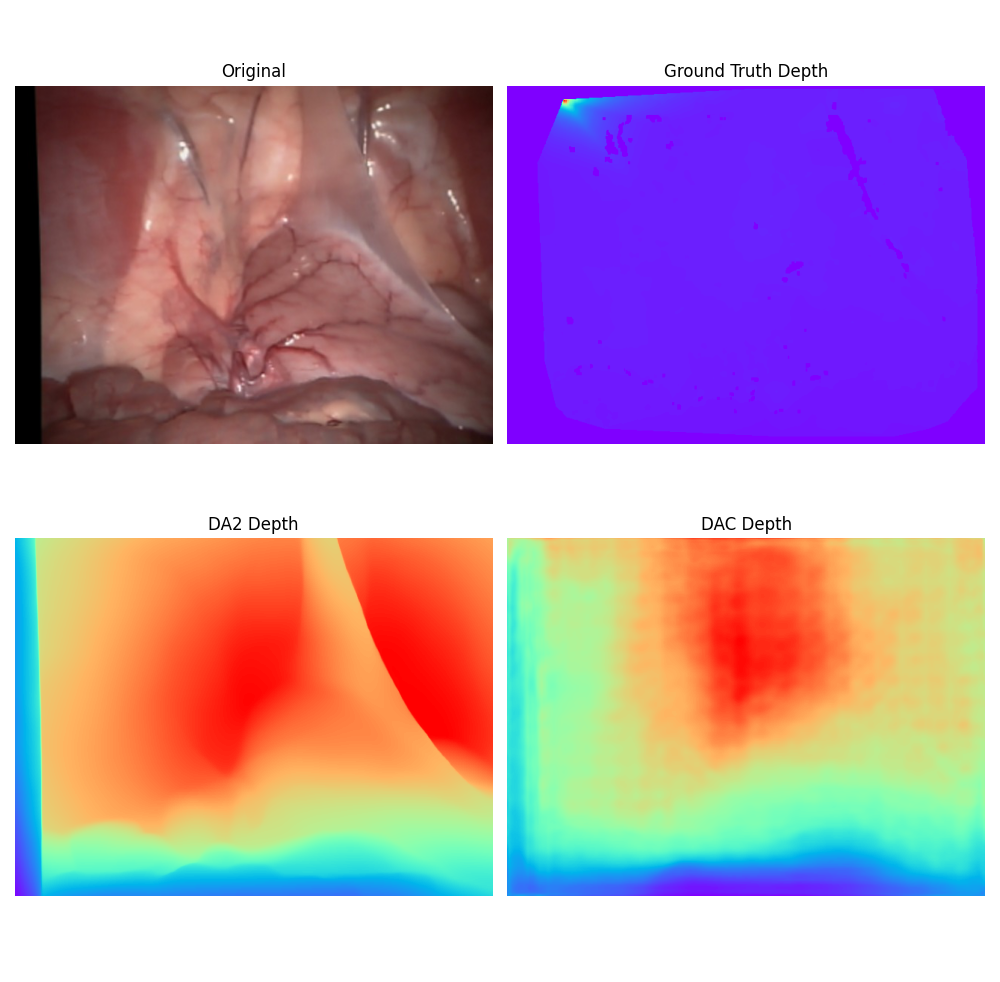}
        \caption{Visualization comparing Depth-Anything V2 and EndoDAC results on a sample frame, turns out to be having an erroneous ground truth.[The 25th frame of rectified from Hamlyn.]}
        \label{fig:errorcomp}
    \end{figure}

\subsection{Searching of the best thresholding scheme}
The seven thresholding schemes are tested on a pair of frames from test22 dataset, using the ground truth depth images, neighbor-ICP method. Different   
\paragraph{Constant Value}
Using a constant value as the threshold.
\[
T = 10
\]

\begin{figure}[H]
    \centering
    \includegraphics[width=0.6\linewidth]{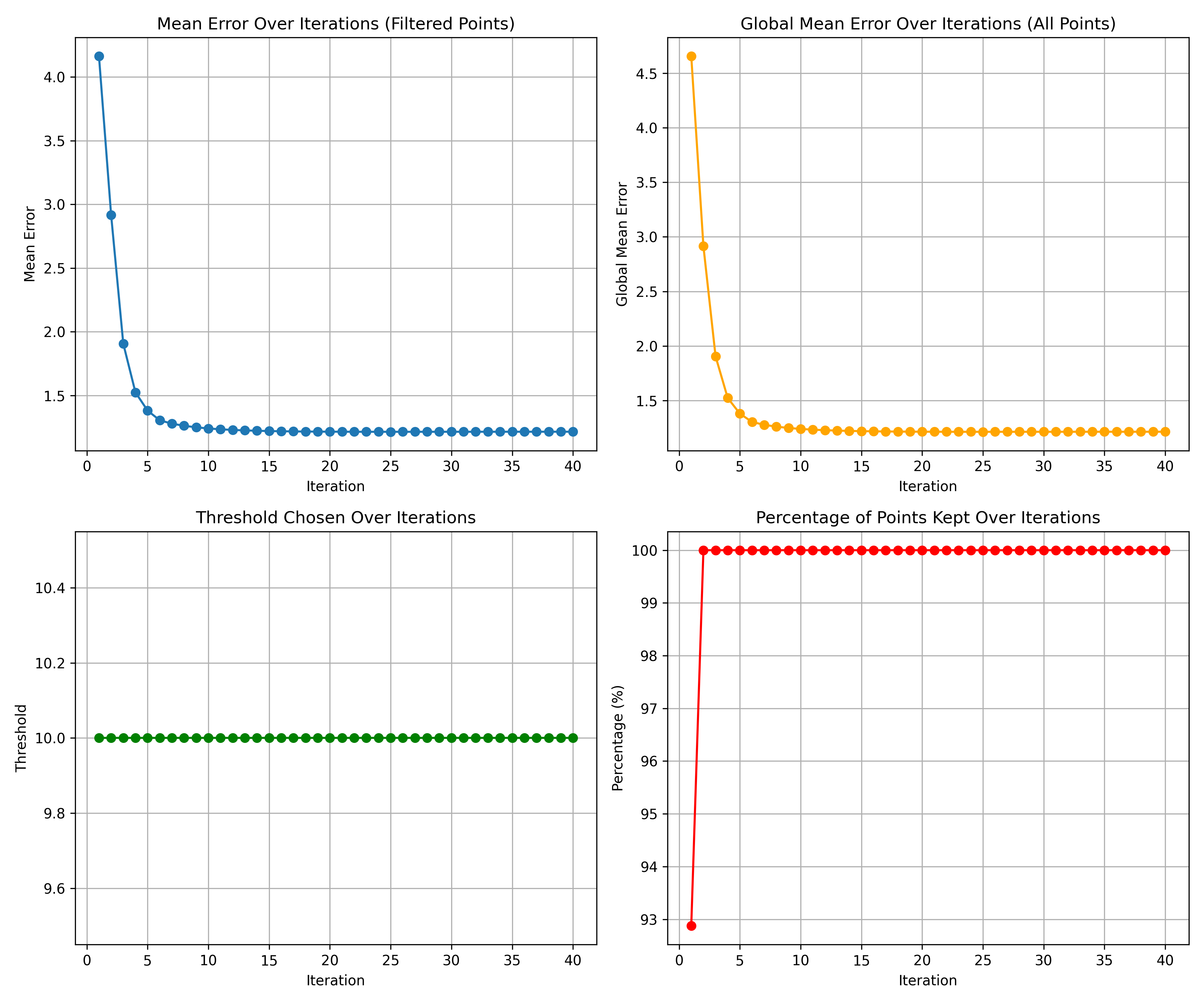}
    \caption{Four plots for constant value method}
    \label{fig:constant}
\end{figure}
\paragraph{90 percentile}
Using the value of 90 percentile of point distance.
\[
T = P_{90}(d)
\]

\text{where } \text{$P_{90}$}(d) \text{ is the 90th percentile of the distances.}
\begin{figure}[H]
    \centering
    \includegraphics[width=0.6\linewidth]{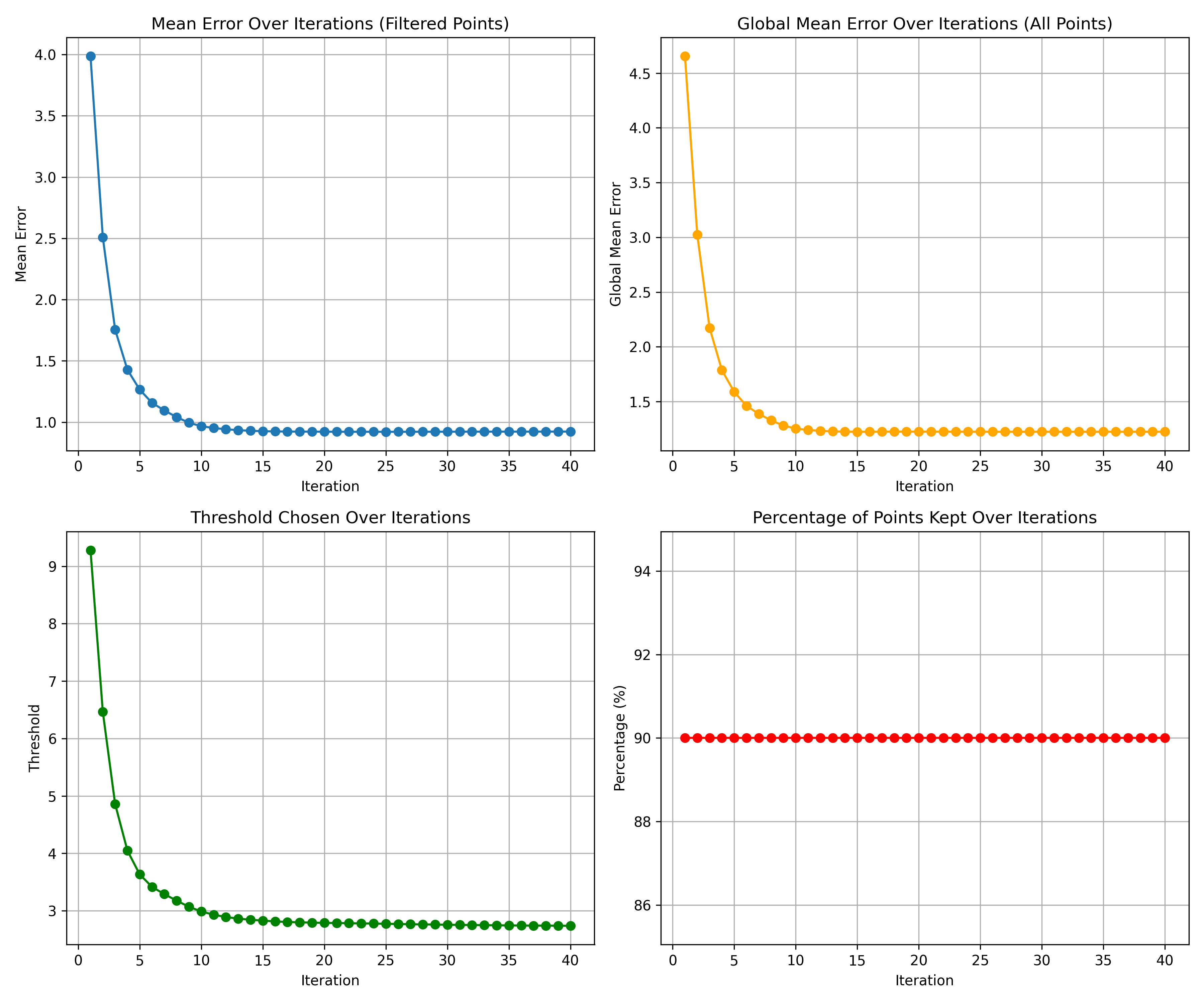}
    \caption{Four plots for 90 percentile method}
    \label{fig:cbtr}
\end{figure}

\paragraph{Mean of error with a factor}
Using the value of a factor multiplied by the mean of distances.
\[
T = 1.5 \times \bar{d}
\]

\begin{figure}[H]
    \centering
    \includegraphics[width=0.6\linewidth]{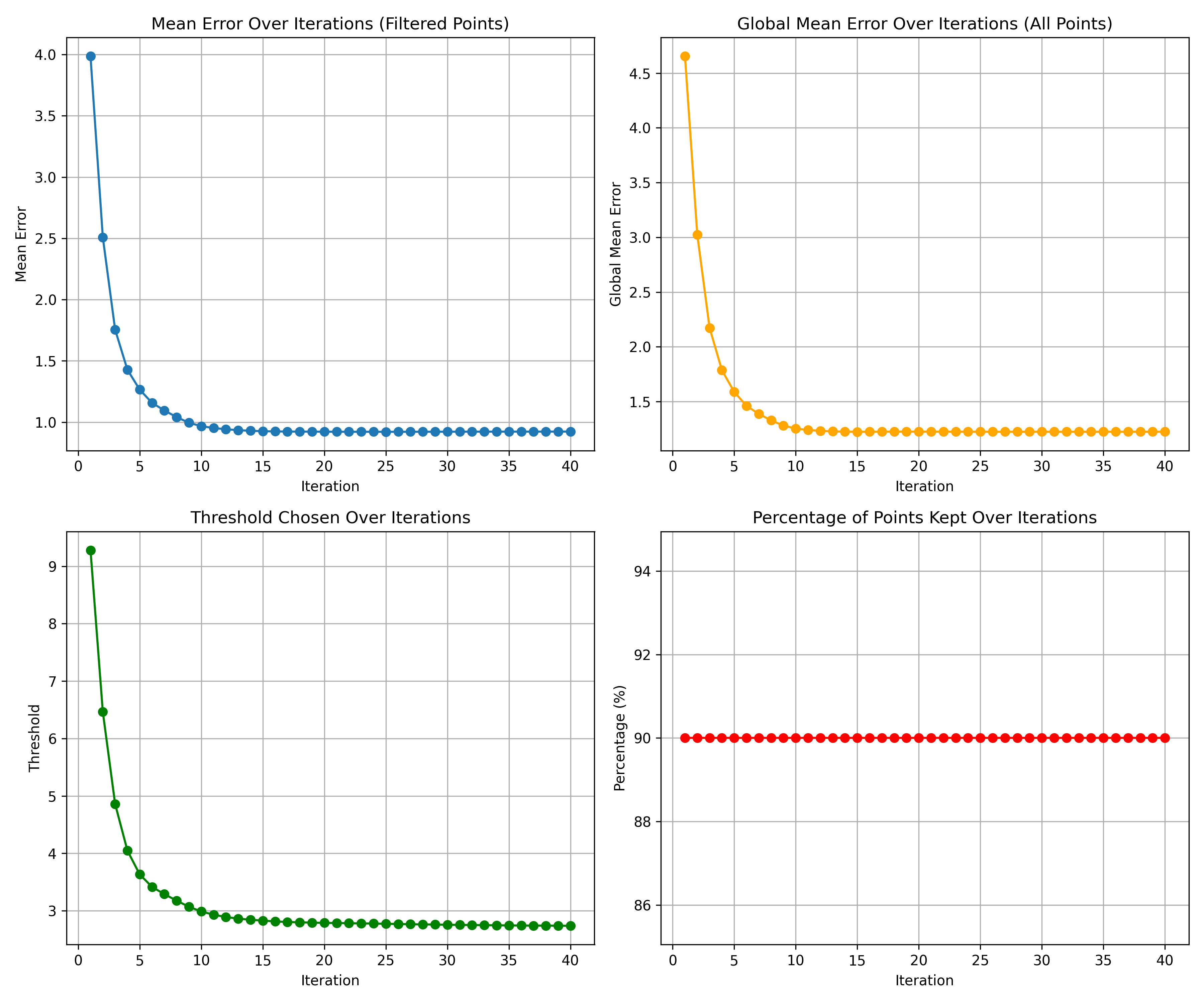}
    \caption{Four plots for mean method}
    \label{fig:cbtr2}
\end{figure}

\paragraph{1.5 Times Median}
Using the value of a factor multiplied by the median of distances.
\[
T = 1.5 \times \text{median}(d)
\]

\begin{figure}[H]
    \centering
    \includegraphics[width=0.6\linewidth]{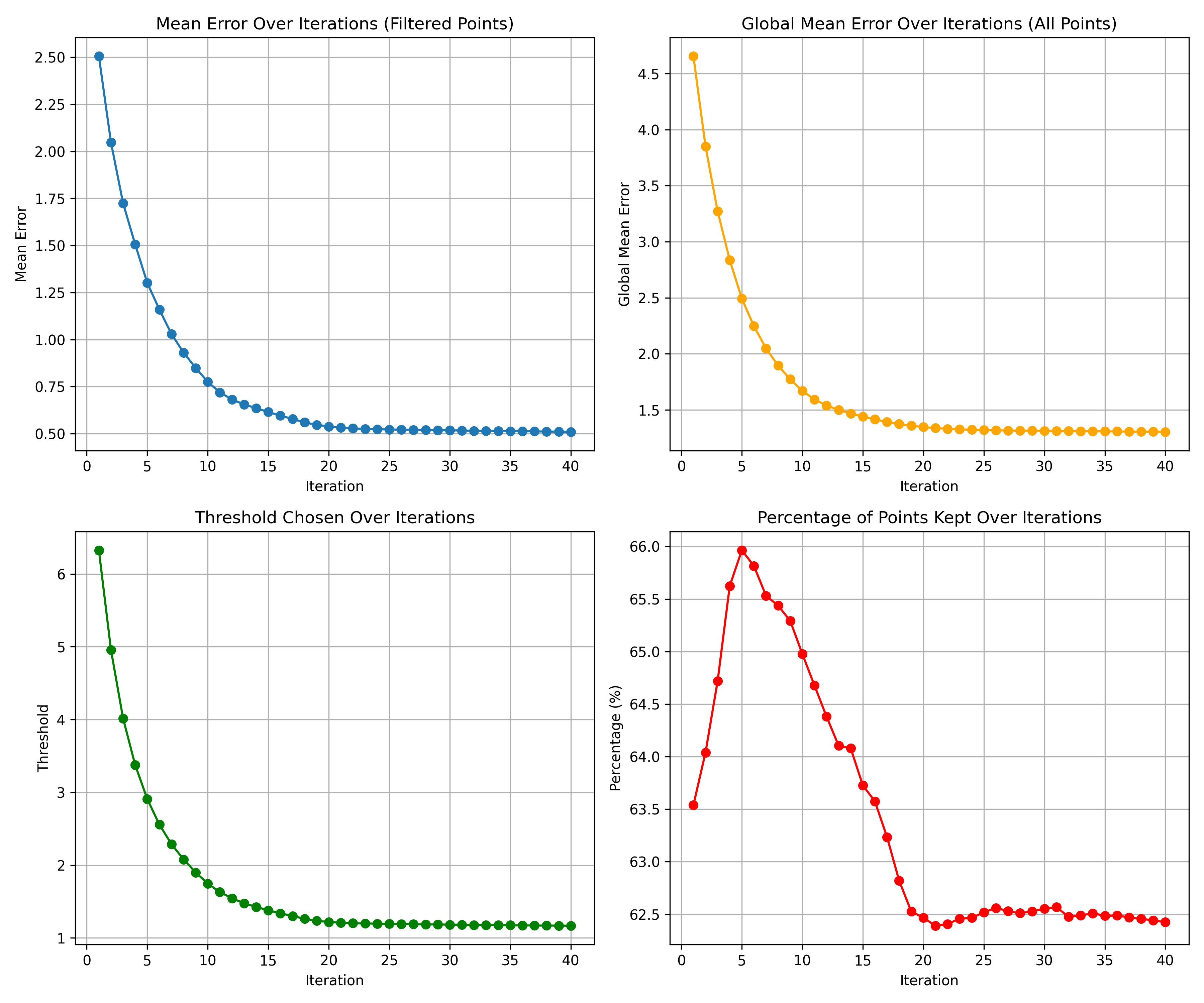}
    \caption{Four plots for 1.5 times median method}
    \label{fig:median}
\end{figure}

\paragraph{Linear Interpolation}
Using linear interpolation between minimum and maximum distances as the threshold. $T_{initial} = 10$$T_{final} = 0.1$
\[
T = a \cdot T_{final} + (1-a) \cdot T_{initial}
\]
\[
a = \frac{i}{max\_iteration - 1}
\]

\begin{figure}[H]
    \centering
    \includegraphics[width=0.6\linewidth]{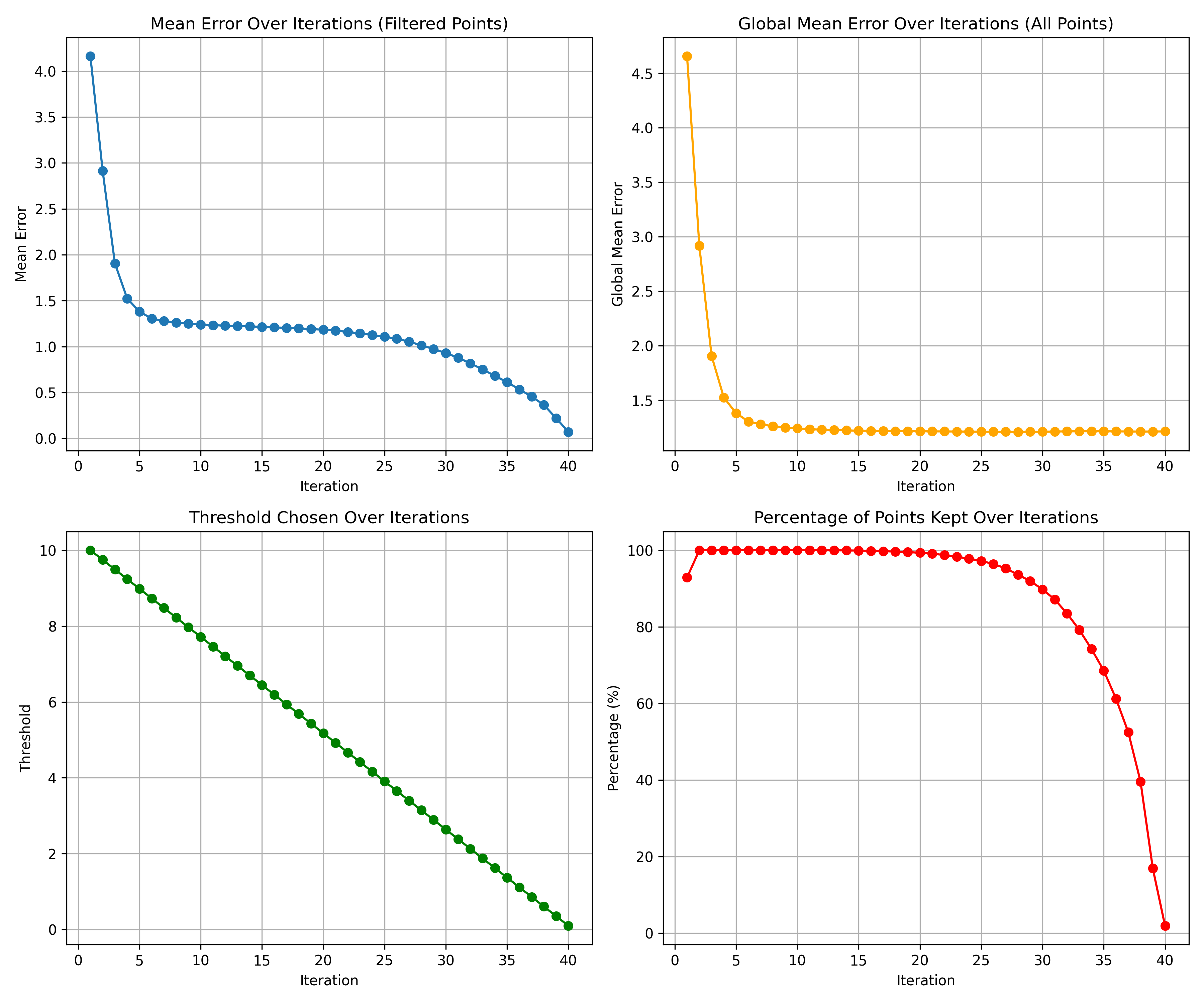}
    \caption{Four plots for linear interpolation method}
    \label{fig:linear}
\end{figure}

\paragraph{80\% of Maximum Distance}
Using a percentage of the maximum distance as the threshold.
\[
T = 0.8 \times d_{\text{max}}
\]

\begin{figure}[H]
    \centering
    \includegraphics[width=0.6\linewidth]{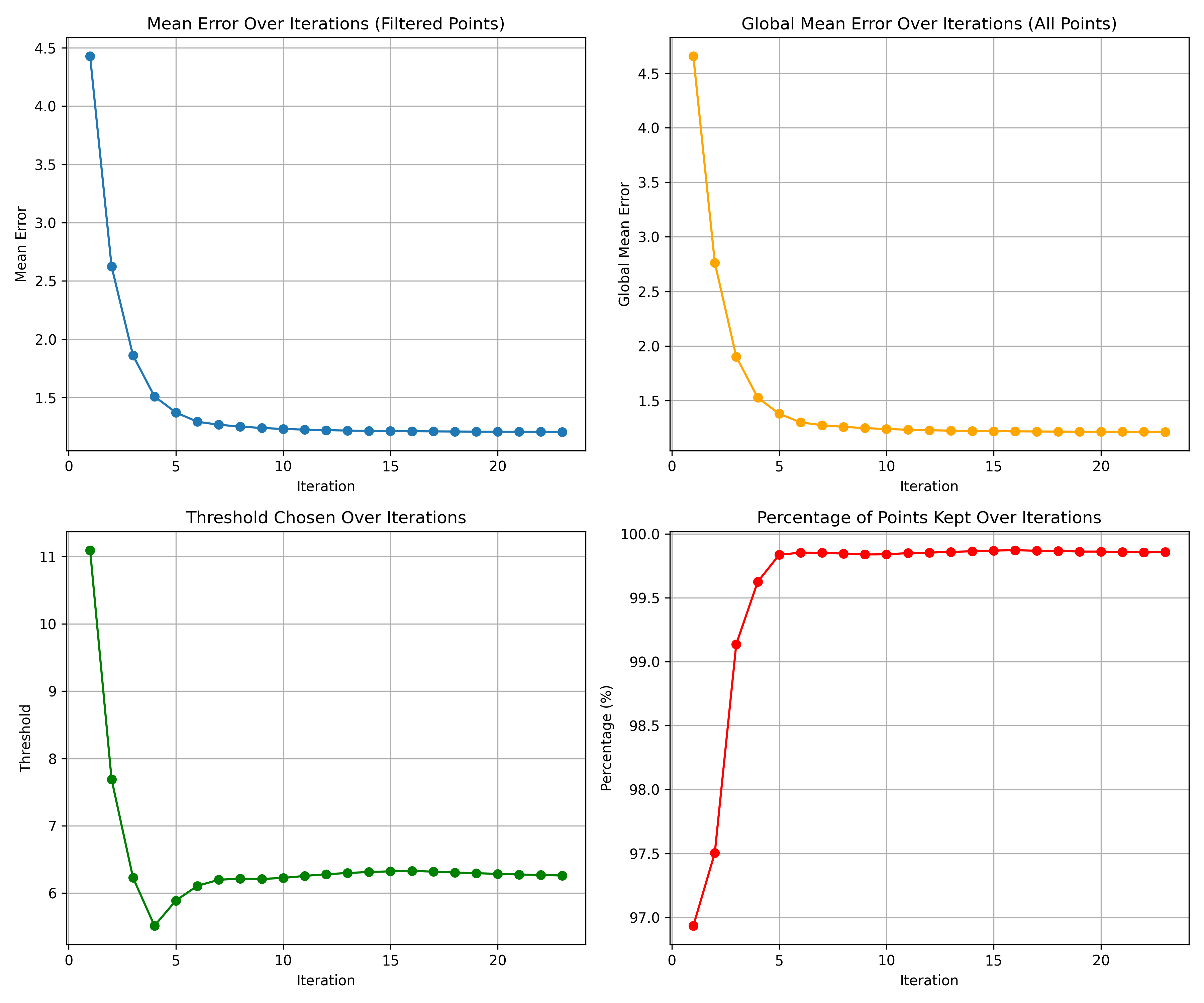}
    \caption{Four plots for 80\% of maximum distance method}
    \label{fig:80pmax}
\end{figure}

\paragraph{Mean Plus Two Standard Deviations}
Using the mean of distances plus two times the standard deviation.
\[
T = \bar{d} + 2 \cdot \sigma_d
\]
\text{where } $\sigma_d$ \text{ is the standard deviation of the distances.}

\begin{figure}[H]
    \centering
    \includegraphics[width=0.6\linewidth]{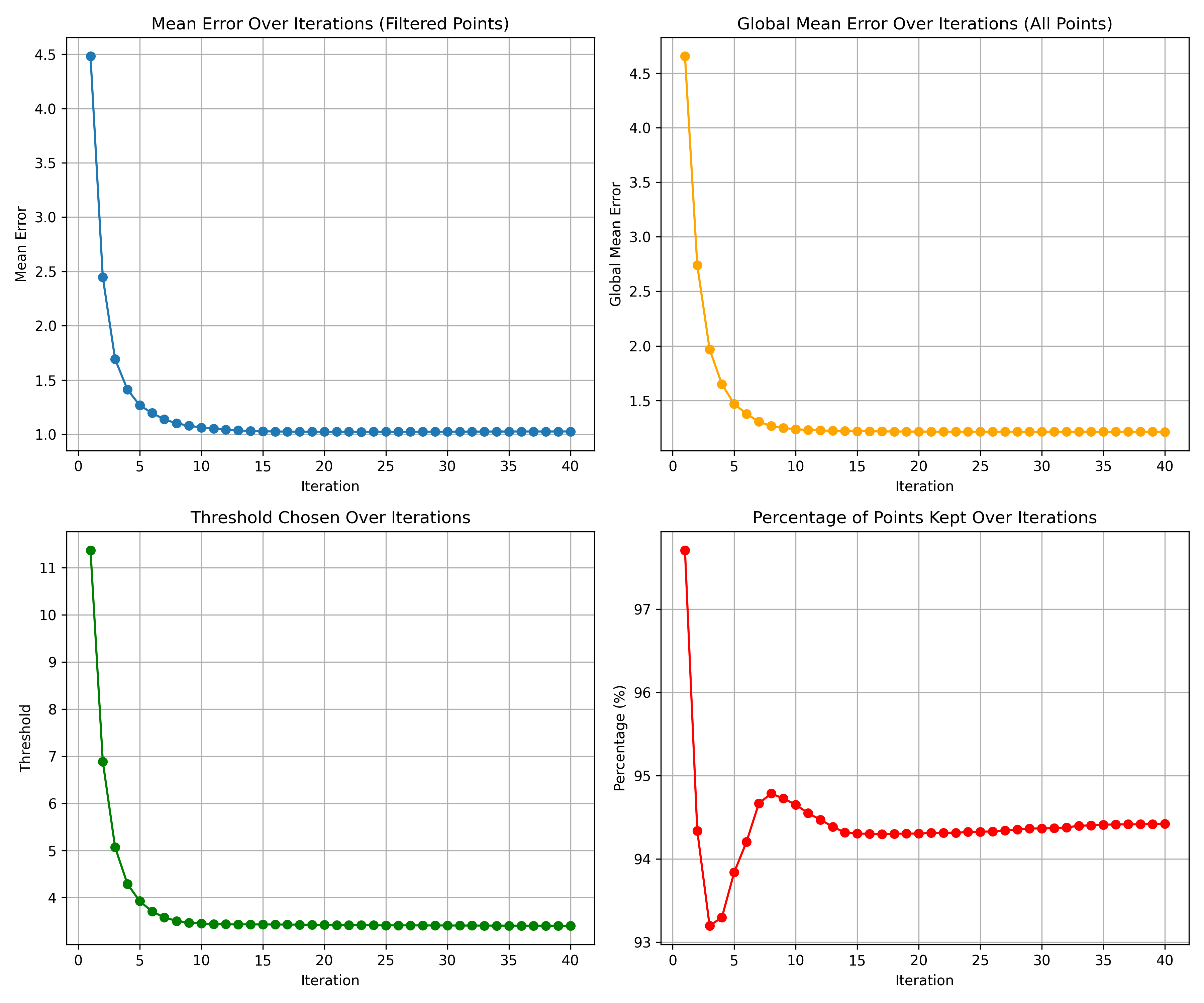}
    \caption{Four plots for mean plus two standard deviations method}
    \label{fig:mean2std}
\end{figure}
\subsection{3D Reconstruction Outcomes}

\paragraph{Qualitative Results}
The outcomes of the 3D reconstruction are recorded and assessed qualitatively. Currently, the scene can be aligned with all points kept, using neighbor-ICP. The best alignment can be found using the mean plus two standard deviation method. A quantitative comparison is not feasible at this point because the discrepancy between the reconstructed scene and the actual scene is too large, and a single scaling is not able to align them. The discrepancy was caused by the nature of the monocular depth estimation since they were relative depth estimation. 

\begin{figure}[H]
    \centering
    \includegraphics[width=0.4\linewidth]{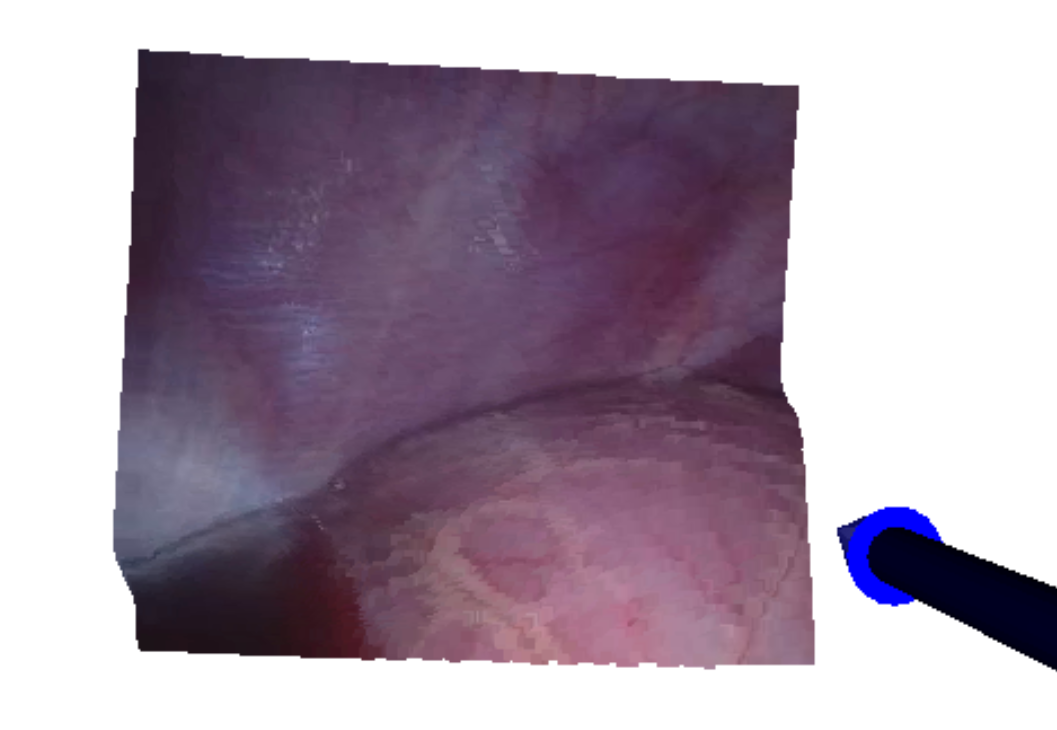}
    \includegraphics[width=0.4\linewidth]{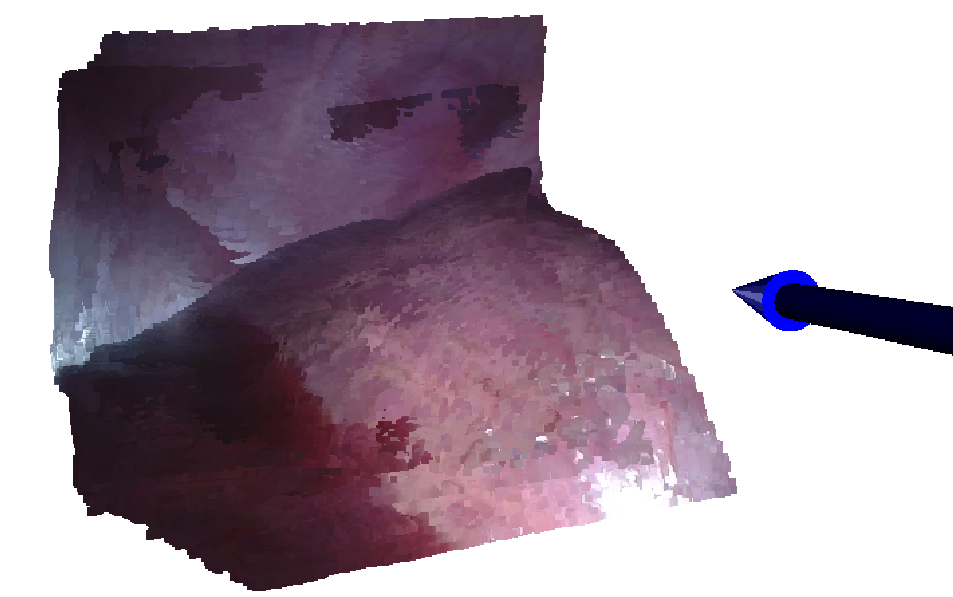}
    \caption{Reconstruction using ground truth depths at frame 1(left) and frame 91(right)}
    \label{fig:recon_result_front}
\end{figure}

\begin{figure}[H]
    \centering
    \includegraphics[width=0.4\linewidth]{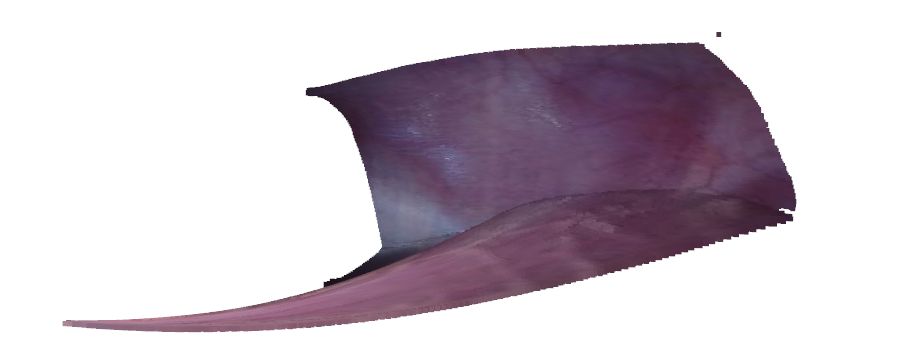}
    \includegraphics[width=0.4\linewidth]{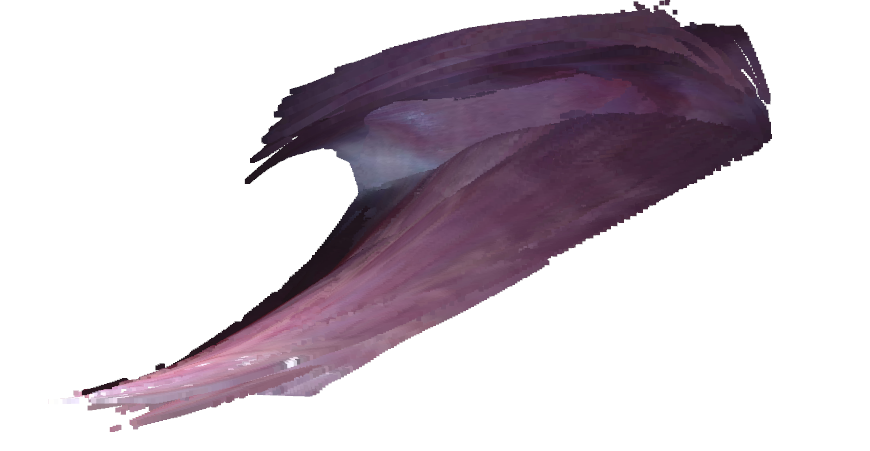}
    \caption{Reconstruction using depth estimation from depth-anything:v2 at frame 1(left) and frame 31(right)}
    \label{fig:recon_result_dav2}
\end{figure}

However, as shown in Figure. \ref{fig:recon_result_front}, the ground truth images are aligned okay, but the depth-estimation images are far worse, see Figure. \ref{fig:recon_result_dav2}

\paragraph{Challenges Faced}
One of the primary challenges stems from the lack of accurate pose estimation, making it difficult to discern which points in sequential frames lie within the shared field of view. Without knowing the camera pose, we cannot reliably identify overlapping regions between successive point clouds, forcing us to rely on heuristic filters. For example, we currently apply a statistical threshold based on the mean and standard deviation of point-to-point distances to differentiate inliers from outliers. This approach helps reduce misalignments, but it often introduces noticeable artifacts, as it cannot distinguish new geometry from previously observed structures.

Additionally, the depth estimation models provide only relative depth values rather than absolute measurements. Ground truth data often have a known depth range (e.g., 100–150 units), but our monocular models produce predictions normalized to a fixed intensity scale (e.g., 0–255). Without a reliable mechanism to align these relative values to a meaningful depth scale, subsequent integration and comparison with ground truth data or between frames remains problematic. This discrepancy underscores the need for improved calibration, fine-tuning of the estimation model, or incorporating additional information (such as stereo or multi-view cues) to achieve more accurate and consistent depth scales.

\subsection{Visualization and Analysis}

\paragraph{ICP Alignment Visualization}
The development of visualization methods allowed for a better understanding of the ICP alignment process. Heatmaps and error plots provided insights into how different thresholding methods affected convergence and accuracy.

\paragraph{Thresholding Strategies}
Dynamic thresholding methods, such as using percentiles or statistical measures, were compared. The experiments suggested that adaptive thresholding could enhance alignment by appropriately balancing the inclusion of corresponding points and the exclusion of outliers.

\section{Discussion}

In this section, we reflect on the observations and experimental insights gained during the pipeline development. Our discussion covers:
\begin{itemize}
    \item the relative-vs-absolute depth challenge
    \item the difficulties encountered when merging point clouds via ICP
    \item the complexities introduced by inconsistent ground-truth data in endoscopic images.
\end{itemize}
\paragraph{Relative Depth}
One of the fundamental challenges in our experiments is that many monocular depth-estimation models (e.g., Depth-Anything:V2 and EndoDAC) produce only relative depth rather than absolute, physically grounded values. Although these relative maps are suitable for certain tasks (like single-frame depth ordering), they introduce ambiguity when performing multi-frame 3D reconstruction. 
Comparing point clouds against ground-truth surfaces requires a consistent metric scale. Since our predicted depths are typically normalized to a range like [0,255], we must attempt a retrospective “best fit” scaling to align them with the ground truth, which leads to additional complexities and potential errors.

Potential solutions include fine-tuning the model with the absolute depth annotations. Also, other zero-shot depth estimation techniques can be implemented, such as the work by Saxena et. al \cite{saxena2023zeroshot}
\paragraph{Reconstruction}
Our experiments showed that thresholding strategies for ICP significantly affect the outcome. A too-lenient threshold incorporates outliers into the alignment, while a too-strict threshold may exclude legitimate correspondences, causing the optimization to converge poorly. Different schemes were tested, and the error over time was plotted. However, each method has trade-offs between speed, stability, and final alignment accuracy.

We found that aligning each new frame to its immediate predecessor (neighbor ICP) sometimes reduces drift compared to aligning directly to a large accumulated global map. However, neighbor-only alignment can compound local errors over a long sequence. Conversely, global alignment can become unstable when the global map is large and contains accumulating errors or artifacts. A hybrid approach, where we periodically “re-anchor” local merges to the global map, may strike a better balance between drift and stability.

\paragraph{Ground Truth Discrepancies}
In our experiments with the Hamlyn dataset, some “ground-truth” frames contained erroneous or incomplete depth annotations. We observed sudden spikes in error metrics such as RMSE or MAE coinciding with suspicious depth maps (Figure. \ref{fig:errorcomp}). This highlights a broader issue: real-world medical datasets can have variable labeling quality, especially if the depth was reconstructed from partial or imperfect data. Thus, before fine-tuning or validating pipeline components, it is essential to identify and exclude frames with invalid ground truth.




\section{Conclusion}
A modular pipeline for frame selection, depth estimation, and 3D reconstruction in endoscopic surgeries has been presented. The system can integrate new algorithms efficiently, as demonstrated by evaluations of Depth-Anything V2, EndoDAC, and multiple ICP thresholding methods. Tests on the Hamlyn dataset show promising results but also highlight core challenges: monocular depth estimations produce only relative scales, while inconsistent ground-truth data complicates alignment. Incorporating pose estimation, refining ICP thresholds, and enhancing data preprocessing is likely to yield more robust reconstructions. The current framework provides a foundation for ongoing research toward real-time, accurate 3D models in clinical endoscopy.

\newpage

\bibliographystyle{ieeetr}
\bibliography{mybib,references}

\end{document}